\documentclass[review]{elsarticle}

\usepackage{lineno,hyperref}
\modulolinenumbers[5]

\usepackage{graphics}
\usepackage{graphicx}
\usepackage[ruled,vlined,linesnumbered]{algorithm2e}

\usepackage{color}
\usepackage{xspace}
\usepackage{amsfonts}
\usepackage{amsmath}

\journal{Discrete Applied Mathematics}

\begin{document}

\begin{frontmatter}

\title{A resource-frugal probabilistic dictionary and applications in bioinformatics}

\author{Camille Marchet, Lolita Lecompte, Antoine Limasset, Lucie Bittner and Pierre Peterlongo}
\address{}




\begin{abstract} 
Indexing massive data sets is extremely expensive for large scale problems. In many fields, huge amounts of data are currently generated, however extracting meaningful information from voluminous data sets, such as computing similarity between elements, is far from being trivial. It remains nonetheless a fundamental need.
This work proposes a probabilistic data structure based on a minimal perfect hash function for indexing large sets of keys.
Our structure out-compete the hash table for construction, query times and for memory usage, in the case of the indexation of a static set. To illustrate the impact of algorithms performances, we provide two applications based on similarity computation between collections of sequences, and for which this calculation is an expensive but required operation.
In particular, we show a practical case in which other bioinformatics tools fail to scale up the tested data set or provide lower recall quality results.


\end{abstract}


\end{frontmatter}


\section{Introduction}
Hardly any research field can escape the current data deluge. 
In particular, genomics produce data volume that is growing extremely rapidly, and will exceed astronomical data in the course of the next decades~\cite{Stephens2015}. Now more than ever, efficient indexing methods are crucial for fully exploiting data. Specifically, the index sizes represent the main limitation constraining the use of high performance clusters fitted out with large RAM facilities. As a consequence, one is often restricted in its computations~\cite{Chen2014,Li2015}.

We propose a novel indexation structure, called ``quasi-dictionary'', that is a probabilistic data structure based on a Minimal Perfect Hash Function (MPHF). It provides a way to associate any kind of data to any input key, scaling up to very large (billions of elements) keys, with a low and controlled false positive rate. By doing so, we focus on the case in which the indexed data are static and huge. 

A number of studies have focused on optimizing non-probabilistic text indexation, using for instance FM-index~\cite{Ferragina2000}, or hash tables. However, except the Bloomier filter~\cite{Charles2008}, to the best of our knowledge, no probabilistic dictionary has yet been proposed for which the false positive or wrong answer rate is mastered and limited. 
The quasi-dictionary mimics the Bloomier filter solution as it enables to associate a value to each indexed element from a static set, and to obtain the value of an element, with a mastered false positive probability if the queried element was not indexed.  Previous studies indicate that the Bloomier filter and the quasi-dictionary have similar execution times~\cite{Charles2008}, whereas our results tend to show that the quasi-dictionary uses approximately ten times less memory. 
Moreover, no available and free Bloomier filter implementation exists so far. 

In the genomic context, the central data represents the content of DNA molecules and is described using sequences in the $\{A,C,G,T\}$ alphabet. 
The indexation is mainly made using $k$-mers, that are words of length $k$, much smaller than the sequences. Basically, $k$-mers are associated to the sequence they belong and are afterwards used for retrieving exact or similar sequences~\cite{altschul1990basic} or for computing inter- or intra-datasets distances~\cite{Benoit2016}.
In the framework of high-throughput sequencing, sequences are called reads and represent short overlapping sub-sequences of genomic material (e.g. chromosomes, RNA) they were sequenced from.


In this work, we use the quasi-dictionary for indexing $k$-mers. This enables to propose two bioinformatics applications described below.
A  key point of these applications is to estimate read 
similarities using $k$-mers diversity only. This so called alignment-free approach is widely used and is a good estimation of similarity measure~\cite{Dubinkina2016}.

Our first application, called \textit{short read connector counter} (SRC\_counter), consists in estimating the number of occurrences of a read (i.e. its abundance) in a read set. This is a central point in high-throughput sequencing studies as abundance is used as an indicator value for reads confidence~\cite{kunin2010trim,Schirmer13012015}. The abundance is also used as a quantitative or semi-quantitative metric to calculate similarity indices between biological communities ~\cite{Amend2010,Kembel2012}.

The second application is called \textit{short read connector linker} (SRC\_linker). It provides a list of similar reads between read sets. Computing read similarity can be performed by a general purpose tool, such as those computing similarities using dynamic programming, and using heuristic tools such as BLAST~\cite{altschul1990basic}. 
However, analyzing all pairs of reads requires a quadratic number of read comparisons, leading to prohibitive computation time, as this is shown in our results section. Tools dedicated to the computation of distances between read sets already exist~\cite{Benoit2016,maillet2014commet,Maillet2012}, but none of them can provide a similarity value for each pair of reads. 
Otherwise, some tools such as starcode~\cite{Zorita2015} are optimized for pairwise sequence comparisons, but, as shown in results, such tools also suffer from quadratic computation time complexity and thus do not scale up data sets composed of numerous reads. We also compared SRC\_linker to alignment-free tools  \cite{li2016minimap, berlin2015assembling,  sovic2016fast} applied on long and highly erroneous read. 
Results show the potential of the proposed approach. As not specialized for long erroneous reads, SRC\_linker show lower precision results than state-of-the-art dedicated tools, but is the only one that combines a scaling up to large problem instances to a high recall and acceptable precision adapting to several error rates.



\paragraph{Availability and license}
Licensed under the GNU Affero General Public License version 3, the quasi-dictionary can be downloaded from \url{http://github.com/pierrepeterlongo/quasi\_dictionary}. 
Our proposed tools SRC\_counter and SRC\_linker were developed using the GATB library~\cite{Drezen2014}. They may be used as stand alone tools or as libraries. They are also licensed under the GNU Affero General Public License version 3 and can be downloaded from \url{http://github.com/GATB/short\_read\_connector}.

\section{Method}
\subsection{Quasi-dictionary index}

\subsubsection{Quasi-dictionary features}
\label{ssec:qdfeatures}
The quasi-dictionary (QD for short) indexes all words from a set $\mathcal{S}$, each over an alphabet $\Sigma$. QD associates each of them to a unique value in $[0,N-1]$, with $N$ being the cardinality of $\mathcal{S}$, denoted by $|\mathcal{S}|$. 
The value QD$[s]$ in $[0,N-1]$, returned for any $s \in \mathcal{S}$, is then used as an pointer to assign any piece of information to $s$. 

Given a quasi-dictionary QD constructed over a set $\mathcal{S}$ and any word $\omega \in \Sigma^*$: \begin{align*}
         \mbox{if } \omega \in \mathcal{S}: & \left\{ \begin{array}{ll}QD[\omega]=i \in [0,N-1] \end{array}
         \right.\\
         \mbox{else:} &  \left\{ \begin{array}{ll} 
         QD[\omega]=i \in [0,N-1] &\mbox{with a probability } p\\
         QD[\omega]=-1 & \mbox{with a probability } 1-p
         \end{array}
         \right.
\end{align*}

Thus, non indexed $\omega \in \Sigma^*$ may be associated to a value in $[0,N-1]$ with a probability $p>0$. This can be seen as a false positive rate. This is why we refer to our index as the \textit{\textbf{quasi}-dictionary}, since it is a probabilistic index. We provide a way to master the $p$ value and to keep it low ($p\approx  2.10^{-4}$).
Note that, querying any indexed $\omega \in \mathcal{S}$ provides a unique and deterministic answer.

\subsubsection{Quasi-dictionary structure}

The core QD indexation structure is a Minimal Perfect Hash Function (MPHF for short) as described in~\cite{Limasset2017}. A MPHF associates any element from a set $\mathcal{S}$ to a unique value in $[0,N-1]$, with $N=|\mathcal{S}|$. A MPHF is not able to detect if a queried element was in the indexed $\mathcal{S}$ set. Thus, in theory, if we had limited the QD structure to a MPHF, then we would have $p=1$. In our context,  this is not totally true, as in practice the MPHF we use may return -1 for a marginally  small subset of elements $\notin \mathcal{S}$. However, using only the MPHF as a data structure would lead to a too high false positive rate. For clarity purpose, for the rest of the paper presentation, we ignore that a MPHF may return -1 for some elements $\notin \mathcal{S}$.

In order to limit the false positive rate $p$, for each indexed element $s \in \mathcal{S}$, we store a fingerprint value associated to $s$, denoted by $fg(s)$, in an array $FG$ of size $N$. Thus $$\forall s \in \mathcal{S}, FG\left[MPHF[s]\right]=fg(s)$$

The fingerprint of a word $s$ is obtained thanks to a hashing function $$ fg:  \Sigma^{|s|} \rightarrow [0,2^f-1]$$ 
A high $f$ value decreases $p$ and increases the memory usage that is $N\cdot f$ bits for the $FG$ array, and \textit{vice versa}.
In practice we chose to use a xor-shift~\cite{marsaglia2003xorshift} hash function for its efficiency in terms of throughput and hash distribution.

Two distinct words have the same fingerprint with a probability $\approx\frac{1}{2^f}$. It follows that there is a probability $\approx\frac{1}{2^f}$ that the quasi-dictionary returns a false positive value despite the fingerprint checking, \textit{i.e.} an $index\neq-1$ for a non indexed word. On the other hand, the $index$ returned for an indexed word is the correct one. In practice we use $f=12$ that limits the false positive rate to $p\approx 2.10^{-4}$. Note that our implementation authorizes any value $f \leq 64$.

\begin{figure}
\centering
    \begin{algorithm}[H]
    \SetAlgoNoEnd
    
    \KwData{ Set $\mathcal{S}$}
    \KwResult{A quasi-dictionary QD indexing elements of $\mathcal{S}$}
    $QD.MPHF = create\_MPHF(\mathcal{S})$ \label{alg:qdcreate:createMPHF}\;
    \ForEach{$\omega$ in $\mathcal{S}$}{ \label{alg:qdcreate:forloop}
        $index = [QD.MPHF[\omega]]$\;
        $QD.FG[index] = fg(w)$\label{alg:qdcreate:createfingerprint}\;
        }
    \textbf{return} QD\;
    
    \caption{Create quasi-dictionary \label{alg:qdcreate}}
  \end{algorithm}
\end{figure}

Finally the QD data structure for a set $\mathcal{S}$ is composed of 
\begin{itemize}
\item the MPHF of elements of $\mathcal{S}$;
\item the $FG$ array composed of $N$ fingerprint values;
\item a user defined array $values$ associating any piece of information to each element of $\mathcal{S}$.
\end{itemize}

\paragraph{QD construction}
Algorithm~\ref{alg:qdcreate} presents the construction of the quasi-dictionary.  The MPHF (line~\ref{alg:qdcreate:createMPHF}) is computed using the MPHF library\footnote{\url{https://github.com/rizkg/BooPHF}, commit number $852cda2$}. This very simple algorithm highlights a drawback of the method: the set $\mathcal{S}$ has to be read twice (line~\ref{alg:qdcreate:createMPHF} and for loop line~\ref{alg:qdcreate:forloop}). Consequently, the QD cannot be created reading data on the fly. Thus the set $\mathcal{S}$ must be precomputed and efficiently readable twice.

\begin{figure}
\centering
    \begin{algorithm}[H]
    \SetAlgoNoEnd
    
    \KwData{Quasi-dictionary $QD$ indexing set $\mathcal{S}$, word $\omega$}
    \KwResult{A value in $[0,|\mathcal{S}|-1]$ or -1 if $\omega$ detected as non indexed}
    $index = QD.MPHF(\omega)$\label{alg:qdqueryMphf}\;
    
    \If{$QD.FG[index] = fg(w)$}{\textbf{return} $index$\;}\label{alg:qdqueryif}
   \textbf{return} $-1$\;
    
    \caption{Query quasi-dictionary \label{alg:qdquery}}
  \end{algorithm}
\end{figure}
\paragraph{Querying the QD}
The querying of a quasi-dictionary with a word $w$ is straightforward, as presented in Algorithm~\ref{alg:qdquery}.

\paragraph{Time and memory complexities}

Our MPHF implementation has the following characteristics.
The structure can be constructed in $O(N)$ time and uses $\approx 3$  bits by elements. We could use parameters limiting memory fingerprint to less than 3 bits per element, but we preferred parameters that allow a great speed up of the MPHF construction and query.
The fingerprint table is constructed in $O(N)$ time, as the $fg$ function runs in $O(1)$. This table uses exactly $N\times f$ bits. Thus the overall quasi-dictionary size, with $f=12$ is $\approx 15$ bits per element. Note that this does not take into account the size of the values associated to each indexed element.

The querying of an element is performed in constant time and does not increase memory usage.

\subsection{Applications of the quasi-dictionary}
In this section we propose applications of our quasi-dictionary data structure for problems that do not meet  scaling methods. There exist several bioinformatic problems that require high scalability, and where scalability is a current bottleneck. In particular we dive into two fundamental bioinformatics applications that can benefit from better indexing structures.

\paragraph{Basic notions and notations} The DNA bases content is read and processed to become large text files trough experiments called sequencing. The sequencing of individual(s) (genomics) or of an environmental  sample (metagenomics) produce read sets. A read set $\mathcal{R}$ is composed of millions or billions of sequences called reads, that are sub-sequence read from the original 
sequenced DNA material. Such sequencing experiments are called High Throughput Sequencing (HTS) because of the subsequent amount of data generated. Note that formally $\mathcal{R}$ should be denoted as a \textit{``collection''} instead of a \textit{``set''} as a read may appear twice or more in $\mathcal{R}$. However, to make the reading easier, we use in this manuscript the term ``\textit{set}'' usually employed for describing HTS outputs. Each read original location on the DNA sequence is unknown. Due to technological limitations, errors during the sequencing occur, generating substitutions (the replacement of a base by another), insertions and deletions (adding or removing one bases or more). The error rate ranges from 0.1\% to 15\% depending on the sequencing technology~\cite{Goodwin2016}. Reads from a set are hopefully redundant. Thus, each locus from a sequenced genome is covered by several reads. The coverage of a sequencing experiment indicates the average coverage of each sequenced genomic locus. 
A DNA sequence can be seen as a sequence written in the {A,C,G,T} alphabet where each represent a molecule that composes the sequence, which we will refer to as base or nucleotide.

Regarding the read set features (redundancy, errors) and sizes, a key treatment is to consider $k$-mers issued from such data sets. A $k$-mer is a word of length $k$ on an alphabet $\Sigma$. Given a read set $\mathcal{R}$, a $k$-mer is said \textit{solid} in $\mathcal{R}$ with respect to a threshold $t$ if its number of occurrences in $\mathcal{R}$ is bigger or equal to $t$. Treatments are performed only on the set of solid $k$-mers.  The choice of $t$ reflects the sequencing coverage and the sequencing quality. This value is usually small, $t=2$ is a common choice: $k$-mers seen only once are likely to be due to sequencing errors and are removed from downstream analyses. 

\paragraph{DNA strands}

DNA molecules are composed of two strands, each one being the reverse complement\footnote{The reverse complement of a DNA sequence is the palindrome of the sequence, in which $A$ and $T$ are swapped and $C$ and $G$ are swapped. For instance the reverse complement of $ACCG$ is $CGGT$.}  of the other. 
As current sequencers usually do not provide the strand of each sequenced read, each indexed or queried $k$-mer should be considered both in the forward and in the reverse complement strand. This is why, in the proposed implementations, we index and query only the canonical representation of each $k$-mer, which is the lexicographically smaller word between a $k$-mer and its reverse complement.

\subsubsection{Short Read Connector Counter}
The first application we propose is called SRC\_counter for \textit{Short Read Connector Counter}. It approximates the number of occurrences of reads from a set $\mathcal{Q}$ in a read set $\mathcal{B}$. 

\begin{figure}[h]
\centering
\noindent\begin{minipage}{\textwidth}
\renewcommand\footnoterule{}                  
    \begin{algorithm}[H]
    \SetAlgoNoEnd
    \KwData{Read set $\mathcal{B}$, read set $\mathcal{Q}$, $k \in \mathbb{N}, t \in \mathbb{N}$}
    \KwResult{For each read from $\mathcal{Q}$, its $k$-mer similarity with set $\mathcal{B}$}
    solid $k$-mer set $\mathcal{R}$ = get\_solid\_kmers$(\mathcal{B},k,t)$ \label{alg:count:createsolid}\;
    quasi-dictionary $QD$ = $create\_quasidictionary(\mathcal{R})$ \label{alg:count:qdcreate}\;
    \ForEach{Solid $k$-mer $\omega$ from $\mathcal{R}$}{
        $QD.values[QD.query(\omega)]=$ number of occurrences of $\omega$ in $\mathcal{B}$\label{alg:count:set}\;
    }
    \ForEach{read $q$ in $\mathcal{Q}$}{
        create an empty vector $count\_q$\;
        \ForEach{$k$-mer $\omega$ in $q$ }{
            \If{$QD.query(\omega) \geq 0$ \label{alg:count:check}}{
                add $QD.values[QD.query(\omega)]$ to $count\_q$ \label{alg:count:get}\;
            }
        }
        Output the $q$ identifier, and (mean, median, min and max values of $count\_q$)\;
    }
    \caption{SRC\_counter: approximating number of occurrences of reads\label{alg:count}}
  \end{algorithm}
  \end{minipage}
\end{figure}

Two  (potentially identical) read sets $\mathcal{B}$ and $\mathcal{Q}$ are considered. The indexation phase works as follows. The solid set of $k$-mers $\mathcal{R}$ from $\mathcal{B}$ is computed (algorithm~\ref{alg:count}, line~\ref{alg:count:createsolid}) using the DSK~\cite{Rizk2013} method. Set $\mathcal{R}$ is indexed (line~\ref{alg:count:qdcreate}) using a quasi-dictionary as presented algorithm~\ref{alg:qdcreate}. The number of occurrences of each solid $k$-mer from $\mathcal{B}$ (line~\ref{alg:count:set}) is obtained from DSK output.

Then starts the query phase. Note that the quasi-dictionary query (line~\ref{alg:count:check}) is performed using algorithm~\ref{alg:qdquery}.  For each read $q$ from set $\mathcal{Q}$, the counts of all its $k$-mers indexed in the quasi-dictionary are recovered and stored in a vector (lines~\ref{alg:count:check} and~\ref{alg:count:get}). Finally, collected counts from $k$-mers from $q$ are used to output an estimation of its abundance in read set $\mathcal{B}$. The abundance is approximated using the mean number of occurrences of $k$-mers from $q$. Median, minimal and maximal number of occurrences of $k$-mers from $q$ are also output. 

\paragraph{Time and Memory complexity}
This algorithm is extremely simple. In addition to the quasi-dictionary creation time and memory complexities, it has a constant memory overhead (8 bits by element in our implementation, limiting maximal counting to 255) and it has an additional $O(\sum_{Q \in\mathcal{Q}}{|Q|})$ time complexity for the query phase. 

\subsubsection{Short Read Connector Linker}\label{soeur}


\begin{figure}[h!]
\noindent\begin{minipage}{\textwidth}
\renewcommand\footnoterule{}                  
    \begin{algorithm}[H]
    \SetAlgoNoEnd
    \KwData{Read set $\mathcal{B}$, read set $\mathcal{Q}$, $k \in \mathbb{N}, t \in \mathbb{N}, f \in \mathbb{N}$}
    \KwResult{For each read from $\mathcal{Q}$, its similarity with each read from set $\mathcal{B}$}
    solid $k$-mer set $\mathcal{R}$ = get\_solid\_kmers$(\mathcal{B},k,t)$ \;\label{alg:link:count}
    quasi-dictionary $QD$ = $create\_quasidictionary(\mathcal{R})$ \label{alg:link:qdcreate}\;

    \ForEach{read $b$ in $\mathcal{B}$}{\label{alg:link:eachb}
        \ForEach{$k$-mer $\omega$ in $b$ }{\label{alg:link:eachkmer}}
            \If{ $QD.query(\omega) \geq 0$ \label{alg:link:check}}{
                add identifier of $b$ to $QD.values[QD.query(\omega)]$ \label{alg:link:addid}\;
            }
        }
    
    \ForEach{read $q$ in $\mathcal{Q}$}{\label{alg:link:startquery}
        create a hash table $targets$\footnote{$target$ keys are read ids, and each $target$ value is a boolean vector of size $|q|$ initially filled with ``$False$''. } \;
        \ForEach{position $i$ in $q$}{
            $\omega$ = $k$-mer occurring position $i$ in $q$\;

            \If{ $QD.query(\omega) \geq 0$ }{
                \ForEach{$tg\_id$ in vector $QD.values[QD.query(\omega)]$}{
                    $targets[tg\_id][i..i+k-1]=``True$''\;\label{alg:link:fillvector}
                }
            }
        }
        Output for $q$ information about positions covered by shared $k$-mers with each\footnote{In practice only reads whose number of positions covered by a shared $k$-mers is higher or equal to a user defined threshold are output} read $tg\_id$ from $\mathcal{B}$. \label{alg:link:out}
            }
    \caption{SRC\_linker: identifying read similarities\label{alg:link}}
  \end{algorithm}
  \end{minipage}
\end{figure}

Our second proposal, called SRC\_linker for \textit{short read connector linker}, compares reads from two (potentially identical) read sets $\mathcal{B}$ and $\mathcal{Q}$. For each read $q$ from $\mathcal{Q}$, a similarity measure with reads from $\mathcal{B}$ is provided. 

The similarity measure we propose for a couple of reads $q\times b$ is the number of positions on $q$ that is covered by at least a $k$-mer that also occurs on $b$. Note that this measure is not symmetrical as the overlapping of shared $k$-mers from $b$ and $q$ may be distinct. More precisely, as reads may be of distinct sizes, the measure can be limited to a window of size $w$ (user defined) on read $q$. The measure is then computed from the  window that maximizes the number of positions covered by shared $k$-mer(s) with $b$.

The indexation phase is described in Algorithm~\ref{alg:link} and  works as follows. A quasi-dictionary is created from solid $k$-mers of set $\mathcal{B}$ (see previous section for comments about lines~\ref{alg:link:count} and~\ref{alg:link:qdcreate}).  Each element of the $QD.values$ table stores for a solid $k$-mer $w$ from $\mathcal{B}$  a list containing the identifiers of reads from $\mathcal{B}$ in which $w$ occurs. See lines~\ref{alg:link:eachb} to~\ref{alg:link:addid}. It is important to notice that $k$-mers $\omega$ considered in the foor loop line~\ref{alg:link:eachkmer} may not belong to $\mathcal{R}$ (non solid $k$-mers) and, in this case, have not been indexed in the quasi-dictionary.


The query phase (lines~\ref{alg:link:startquery} to the end) works as follows. Each $k$-mer of each input query read $q$ is queried in the quasi-dictionary. If this $k$-mer is associated to one or several reads from $\mathcal{B}$, then for each of them, one reminds using a boolean vectors, positions on $q$ covered by this shared $k$-mer.

Once all $k$-mers of a read $q$ are treated, the identifier of $q$ is output and for each read $b_j$ from $\mathcal{B}$ its identifier is output together with the number of shared $k$-mers with $q$ (line~\ref{alg:link:out}). In practice, in order to avoid quadratic output size and to focus only on similar reads, only reads sharing a number of $k$-mers higher or equal to a user defined threshold are output. 
Note that, for clarity purpose, Algorithm~\ref{alg:link} does not detail in line~\ref{alg:link:out} how the best window of size $w$ is selected from each target read. This step is straightforward. It simply detects for each targeted $tg\_id$ reads the $i$ value such as $targets[tg\_id][i..i+w-1]$ maximizes the number of $True$ values.

\paragraph{Time and Memory complexity}
In addition to the quasi-dictionary data structure creation, considering a fixed read size, Algorithm~\ref{alg:link} has  $O(N\times \overline{m})$ memory complexity and a $O(N+\sum_{Q \in\mathcal{Q}}{|Q|\times \overline{m}})$ time complexity, with $\overline{m}$ the average number of distinct reads from $\mathcal{B}$ in which a $k$-mer from $\mathcal{Q}$ occurs. In the worst case $\overline{m}=N$, for instance with $\mathcal{B}=\mathcal{Q}=\left\{A^{|\text{read}|}\right\}^N$. In practice, in our tests as well as for real sets composed of hundred of million reads, $\overline{m}$ is limited to $\approx 2.22$. 

\section{Results}
This section aims at presenting the quasi dictionary data structure performances and its applications results in the context of bioinformatics.
In the first part of this section are presented the general performances of the quasi-dictionary in comparison to a broadly used hash table. We also use the SRC\_counter results to show the practical impact of false positive calls when using the quasi-dictionary.

In the second and third parts, we focus on the performances of the quasi-dictionary application SRC\_linker. This reflects the global and fundamental need in bioinformatics for tools able to retrieve similar reads between to read sets. Current technologies enable the access to two main different types of reads: short and long. Each read type brings its specific algorithmic problems. Short reads comes in more voluminous data sets. Long reads experiments contain usually less reads but longer sequences and more $k$-mers. They also have a subsequently higher error rate than small reads. Their noisy nature makes it difficult to obtain a good sensitivity. Each type of reads, short and long, implies the direct need of scaling methods and has its own dedicated tools that adapt to its specificity. This explains why tools in the benchmarks of the second section differ from tools in the third section. This is thus the occasion to demonstrate SRC\_linker's successful adaptability to both read types. 

In the second part (Section~\ref{ssec:similar_short}), we focus on short reads similarity computation. We compare to state of the art alignment tools to highlight the gain in scaling we provide.

In the third part (Section~\ref{ssec:long} ), we tackle the long reads problem and show that our lightweight data structure enables us to be more robust to the high error rate than most of the tools and to be more scalable than another robust tool.

All tests performed on short reads (Sections~\ref{ssec:qdperf} and~\ref{ssec:similar_short}) from a metagenomic \textit{Tara} Oceans~\cite{Karsenti2011} read set ERR59928\footnote{ \url{http://www.ebi.ac.uk/ena/data/view/ERR599280}}, composed of 189,207,003 reads of average size 97 nucleotides.  From this read set, we created sub-sets by selecting first 100K, 1M, 10M, 50M and 100M reads (with K meaning thousand and M meaning million). Data set used for long reads experiments is described Section~\ref{ssec:long} and in the Appendix.

Tests were performed on a Linux 20-CPU nodes running at 2.60 GHz with an overall of 252 GBytes memory.

\subsection{Quasi-dictionary performances}
\label{ssec:qdperf}

We first performed tests enabling the evaluation of the gain of our proposed data structure when compared to a classical hash table. Secondly we provide results that enable to estimate the impact of false positives.

\subsubsection{Standard hash table compared to quasi-dictionary index}

\begin{table}[h]
\centering
\begin{tabular}{c|cc|ccc|cc}
 \begin{tabular}{c}Indexed Data set\\(nb indexed $k$-mers)\end{tabular}& \multicolumn{2}{c|}{\begin{tabular}{c}
  Construc.\\ time (s)
 \end{tabular}} &  \multicolumn{3}{c|}{Memory (GB)}&\multicolumn{2}{c}{\begin{tabular}{c}Query\\ Time(s)\end{tabular}}\\
   &QD  &Hash & QD & QD62 & Hash & QD  & Hash\\ \hline
 1M (64$\times 10^6$) &  16 & 96 & 0.23 & 0.61  & 2.46 & 11 & 17\\
 10M (622$\times 10^6$) & 174 & 979 & 1.78 & 5.40 & 23.58 & 11 & 17\\
 50M (2,813$\times 10^6$) & 538 & 4,445 & 7.92 & 24.29 & 106.23 & 11 & 19 \\
 100M (5,191$\times 10^6$) & 1,322 & 7,995 & 14.58 & 44.80 & 202.88 & 13 & 19\\
 Full (8,784$\times 10^6$) & 2,649 & - &  24.75 & 75.88 & - & 15 & -\\
\end{tabular}
\caption{
Wallclock time and memory used for creating and for querying the quasi-dictionary using the default fingerprint size $f=12$ (denoted by ``QD'') and the C++ \textit{unordered\_map}, denoted by ``Hash''. Column ``$k$-mer count time'' indicates the time DSK spent counting $k$-mers. 
Tests were performed using $k=31$ and $t=1$ (all $k$-mers are solid). The query read set was always the 10M set. We additionally provide memory results using the quasi-dictionary with a fingerprint size $f=62$ (denoted by ``QD62''). Construction and query time for QD62 are not shown as they are almost identical to the QD ones. On the full data set, using a classical hash table, the memory exceeded the maximal authorized machine limits (252 GB).}
\label{tab:qdbench}
\end{table}

We tested the quasi-dictionary performances by indexing iteratively the  read subsets plus the full ERR59928 set, each time querying reads from set 10M. We compared our solution performances with a classical indexation scheme done using the C++11 \textit{unordered\_map} hash table. Results are presented in Table~\ref{tab:qdbench}. These results show that the quasi-dictionary is roughly an order of magnitude faster than a classical hash table solution. Moreover, the quasi-dictionary memory footprint is in average $10$ times smaller. These results show that the hash table is not a viable solution scaling up current read sets composed of several billions $k$-mers. Notably, using a hash table, the full data set could not be indexed  because of memory limits.
Results also highlight the fact that the query is fast and only slightly depends on the number of indexed elements.

Importantly, using a fingerprint large enough, we can force the quasi-dictionary to avoid false positives. 
Here we used $f=62$, so any 31-mer on the alphabet $\{A,C,G,T\}$ can be assigned to a unique value in $[0,2^{62}-1]$ and \textit{vice versa}.
As expected, the quasi-dictionary data structure size increases when $f$ increases. Interestingly, on this example and as shown in Table~\ref{tab:qdbench}, the size of the quasi-dictionary with $f=62$ remains on average 4 times smaller than the size of the hash-table. Keeping in mind that the quasi-dictionary is faster to construct and to query, the usage of this data structure avoiding false positives presents only advantages compared to the hash table usage for indexing a static set. However, we recall that avoiding false positives works well here because one an alphabet of size four, using a fingerprint of size $f=62$ is sufficient to represent exactly any 31-mer.
With larger alphabets such as the amino-acids (22 characters) or the Latin ones, the usage of a hash table is recommended if false positives are not tolerated.

\subsubsection{Approximating false positives impact}
We propose an experiment to assess the impact on result quality when using a probabilistic data structure instead of a deterministic one for estimating read abundances. Thus in this experiment, we used the SRC\_Counter tool, used for estimating the read coverage of a read set. 

We used the 100M reads set both for the indexation and the querying, thus providing an estimation of the abundance of each read in its own read set. 
We made the indexation using $k=31$ and $c=2$. Note that, with $c=2$ only $k$-mers seen twice or more in the set are solid and indexed. In this example only 756,804,245 $k$-mers are solid among the 5,191,190,377 distinct $k$-mers present in the read set. This means that 85.4\% of queried $k$-mers are not indexed, this matter of fact enables to measure the impact of the quasi-dictionary false positives. 
We applied the SRC\_Counter algorithm as described in Algorithm~\ref{alg:count}, using $f=12$ or $f=62$. With $f=12$, the false positive rate is non null (see Section~\ref{ssec:qdfeatures}) while with $f=62$, as previously mentioned, there are no false positives with $k=31$. These two experiments thus enable to evaluate the impact of false positives when using the quasi-dictionary for downstream analyses such as read abundance estimations. 

Because of the quasi-dictionary false positives, counts obtained  with $f=12$ are an over-estimation of the real result obtained with $f=62$. Thus, we computed for each read the observed difference in the counts between results obtained with the two approaches. 
The max over-approximation is 26.9, and the mean observed over-approximation is $7.27\times10^{-3}$ with a $3.59\times10^{-3}$ standard deviation. Thus, bearing in mind that the average estimated abundance of each read is $\approx 2.22$, the average count over-estimation represents $\approx0.033\%$ of this value. Such divergences are negligible for downstream analyses.


\subsection{Identifying similar reads on short sequences}
\label{ssec:similar_short}
This section deals with short reads experiments that need highly scalable structures because of the voluminous data sets often encountered.
We set a benchmark of the SRC\_linker method with comparisons to state of the art tools that can be used in current pipelines for the read similarity identification presented in this paper.
We compared our tool using the default parameters with the classical method 
BLAST~\cite{altschul1990basic} (version 2.3.0), with default parameters. BLAST is able to index big data sets, and consumes a reasonable quantity of memory, but the throughput of the tool is relatively low and only small data sets were treated within the timeout (10h, wallclock time).
We also included two broadly used mappers, that may be used for finding read similarities.
We used Bowtie2~\cite{langmead2012fast} (version 2.2.7), and BWA~\cite{li2009fast} (version 0.7.10). By default these two tools only output the best possible alignment found. To enable the comparison with BLAST and our method, we used the ``any alignment'' mode (\texttt{-a} mode in Bowtie2, \texttt{-N} for BWA) in order to output all alignments found instead of the best one only. Both tools are not well suited to index large set of short sequences nor to find all alignments and therefore use considerably more resources than their standard usage. We also compared SRC\_linker to starcode (1.0), that clusters DNA sequences by finding all sequences pairs below a  distance metric\footnote{The Levenshtein distance defined as the minimum number of insertions, deletions and substitutions needed to transform one sequence into another}. 

Importantly, notice that such comparisons are unfair insofar as compared tools provides much more precise distance information between pair of reads than SRC\_linker and performs additional tasks. However, our benchmark highlights the fact that such approaches suffer from intractable number of read comparisons, as demonstrated by presented results. 

\begin{table}[h]
\centering

\begin{tabular}{c|c|c|c|c|c|c}
 & Data set &  100K & 1M & 10M & 100M & Full\\
 & \begin{tabular}{c}Nb solid\\$k$-mers ($\times 10^6$)\end{tabular} & 0.2 & 0.6 & 22 & 757 & 1,880\\
 
 \hline\hline
     & Blast &  52 & 795 & - & - & -\\
     & Bowtie2   & 51 & 10,644 & - & - & -\\
Time& BWA       & 106 & 3,155 & 62,912 & - & - \\
 (s)     & starcode  & 29 & 1,103 & 131,139 &- & - \\
     & SRC\_linker  & \textbf{5} & \textbf{45}  & \textbf{587} & \textbf{14,748} & \textbf{40,828} \\
     \hline
     & Blast  & 18.5 & 24.5 & - &- & - \\
     & Bowtie2   & 0.77 & 5.54 & - &  -& -\\
Memory& BWA       & \textbf{0.49} & 3.4 & 5.9 &- &  -\\
   (GB)    & starcode  & 12.06 & 18.18 & 73.5 & -&  -\\
     & SRC\_linker  & 1.07 &\textbf{ 1.28} & \textbf{3.61} & \textbf{44.37} & \textbf{110.84} \\
    
\end{tabular}

\caption{CPU time and memory consumption for indexing and querying a data set versus itself. Tests were performed using $k=31$ and $t=2$ ($k$-mers seen twice or more are solid). We set a timeout of 10h. BLAST crashed for 10M data set, Bowtie2 reached the timeout we set with more than 200h (CPU) for 10M reads. BWA reached the timeout for 100M reads (more than 200h (CPU) on this data set). On the 100M data set, starcode also reached the timeout. Only SRC\_linker finished on all data sets. }
\label{tab:mappersbench}
\end{table}

Because of the time or memory limitations we could compare against all methods only up to 1M reads. Results are reported Table~\ref{tab:mappersbench}. BWA performed better than the two other tools in terms of memory, being able to scale up to 10M reads, while Bowtie2 and BLAST could only reach 1M reads comparison. On this modest size of read set, we see that SRC\_linker is already ahead both in terms of memory and computation time. The gap between our approach and others  increases with the amount of data to process. Dealing with the full data set reveals the specificity of our approach, being the unique able to scale such data set.

\subsection{Using SRC\_linker on long, spurious sequences}
\label{ssec:long}
In this section we focus on long, spurious reads. They appeared in the last few years and are longer at least an order of magnitude from short reads (thousands of bases instead of hundreds for short reads technologies). They notably come at the price of a highly increased error rate: up to 15\% \cite{laehnemann2016denoising} and even get higher \cite{jain2015improved,ip2015minion} depending on the technology used, while it is rather lower than 1\% in short reads. This time, scaling challenges are intertwined with sensitivity challenges as the sequences are very noisy.


A vast majority of $k$-mers created by sequencing errors in the sequences do not  exist in the original DNA. Moreover as we look for small $k$-mers, depending on the coverage and on the size of $k$-mers, it can be likely that many spurious $k$-mers show occurrences above 2 in the data set \cite{carvalho2016improved}. This states that having a good recall while remaining precise is a real hard task. We remind that the recall represents the number of relevant element retrieved among all relevant elements. 
The precision represents the number of relevant element among the retrieved elements. 
The recall and precision formulas used in this framework are proposed in the appendix.

Relying on the quasi-dictionary, we argue we can afford to index all (solid) $k$-mers at a reasonable cost and then benefit from a more complete information about the content of the reads. This has a positive effect on our recall. We also controlled that our precision would remain high.
Since these technologies are only a few years old, a small number of tools exist in the literature to handle the long reads in terms of retrieving similarities. To tackle these issues, other state of the art tools like Minimap \cite{li2016minimap} and MHAP \cite{berlin2015assembling} make the choice to index only a subset of $k$-mers using minimizers~\cite{roberts2004reducing}. Another tool, GraphMap \cite{sovic2016fast} does not use minimizers but relaxes the condition of exact matches of $k$-mers, by using seed designs that allow errors. 


In order to estimate the impact of indexing all $k$-mers using the quasi-dictionary in the context of long spurious reads comparisons, we simulated long reads coming from different  regions of a genome, and we compared the three described approaches to ours. 
Our goal here is to demonstrate the potential offered by the quasi-dictionary data structure. 
Thus we must precise that  the comparison proposed is not absolutely fair to the extent that the four presented tools do not make the same choices once reads are mapped. MHAP and GraphMap offer post-treatments while Minimap and SRC\_linker stick to the single recruitment phase. 
Versions of each software, command lines, and the detailed description of the simulation are provided in Appendix.

\begin{table}[h]
\centering
\begin{tabular}{c|c|c|c|c}
                  & Minimap      & MHAP      & GraphMap          & SRC\ linker\\ \hline

Recall(\%)             & 99.31 & 86.54 & 97.77 & 97.96\\
Precision(\%)          & 99.83 & 96.74 & 99.53 & 99.58\\ 
F-measure              & 99.57 & 91.35 & 99.15 & 98.76\\
\hline \hline

Memory (GB)        & 6.37 & 25.94 & 11.87 & 2.55\\
Time (m:ss)        & 0:24 & 3:19 & 9:00 & 5:49 \\ 
\end{tabular}
\caption{Precision and recall followed by time and memory performances for 100K simulated long reads on 1K distinct regions on the C. elegans genome, with 12\% error rate. The F-measure is the harmonic mean of precision and recall.}
\label{aln-free}
\end{table}




\begin{table}[h]
\centering
\begin{tabular}{c|c|c|c|c}
                  & Minimap & MHAP & GraphMap & SRC\_linker \\ \hline

Recall(\%)          & 63.25 & 14.21 & 92.08 & 91.95 \\
Precision(\%)       & 99.89 & 86.94 & 99.72 & 97.89 \\ 
F-measure           & 77.46 & 24.43 & 95.75 & 94.83\\
\hline \hline

Memory (GB)         & 6.38 & 26.32 & 12.03 & 2.65 \\
Time (m:ss)         & 0:24 & 3:20 & 9:09 & 7:35\\
\end{tabular}
\caption{Precision and recall followed by time and memory performances for 100K simulated long reads on 1K distinct regions on the C. elegans genome with a 15\% error rate.}
\label{aln-free2}
\end{table}

The results from two simulations of increasing error rates are presented in Tables~\ref{aln-free} and~\ref{aln-free2}. 

From those results we can see that the key advantage expected from minimizers methods, time and memory low footprint, is met with Minimap and only half-met with MHAP (which shows a quite high memory consumption). If SRC\_linker running time if quite high and comparable to GraphMap's, it however has the lowest memory footprint, while indexing more elements than Minimap and MHAP. 

On 100K long reads with 12\% error (Table~\ref{aln-free}), with the exception of MHAP recall, all tools present near perfect precision and recall.
All three state of the art tool provide a high precision rate on this first experiment, as expected provided the filters they embed. 
As we increased the error to a more difficult scenario (Table~\ref{aln-free2}), we can see that MHAP and Minimap recall scores decrease while GraphMap maintains very high recall and precision. Our recall outperforms those of other tools, however, as expected, we reach a lower precision than GraphMap.

This shows that SRC\_linker already provides acceptable precision without any post-treatment on the contrary to GraphMap and MHAP that use downstream filters. 
This also shows that we successfully mimic GraphMap in its ability to adapt to varying error rates.


Minimap is presented as an experimental tool and has not the ambition to reach the recalls of the other tools that integrate more developments. Its force relies in its lightweight and fast execution. A third data set of higher size (1M reads for 10K distinct regions) is generated to show the scalability of each method. GraphMap ran for more than 15 hours on the bigger data set thus reached the timeout we set. MHAP crashed on this bigger data set. All in all only Minimap and SRC\_linker managed to scale on the bigger data volume. 
They obtained following results: 98.56\% recall and 97.95\% precision for Minimap, and 98.28\% recall and 92.63\% precision for SRC\_linker.

In these experiments we chosen the parameters of SRC\_linker to optimize its F-measure, giving results not always in favour of precision. The SRC\_linker precision could be improved using downstream filters or more stringent parameters. 

We shown that the tool we provide, while being simple works well as a recruitment tool for highly noisy sequences thanks to its ability to preserve as much information as possible about the sequences. It presents both advantages to be robust to changes over errors or read length, and to be scalable.

Importantly, we recall that our message here is not to outperform state-of-the-art long read mappers that were designed and optimized specifically for this task. We simply want to show the application potential that our data structure offers.

\section{Conclusion}

In this contribution, we propose a new indexation scheme based on a Minimal Perfect Hash Function (MPHF) together with a fingerprint value associated to each indexed element. 
Our proposal is a probabilistic data structure that has similar features than Bloomier filters, with smaller memory fingerprint.
This solution is resource-frugal (we have shown experiments on sets containing more than eight billion elements indexed in $\approx 44$ minutes and using less than 25GB RAM) and opens the way to new (meta)genomic applications. As proofs of concept, we proposed two novel applications: SRC\_counter and SRC\_linker. The first estimates the abundance of a sequence in a read set. The second detects similarities between pair of reads inter or intra-read sets. These applications are a start for broader uses and purposes.

Two main limitations of our proposal due to the nature of the data structure can be pointed out.
Firstly, compared to standard hash tables, our indexing data structure presents the following drawback: the exact set of keys to index has to be static and defined previously to the data structure creation. 
This is a clear limitation for non fixed set of keys. Secondly, our data structure can generate false positives during query. Even with the proposed false positive ratio limited to $\approx 2.10^{-4}$ with defaults parameters, it may not be adapted for all applications. However we can force our tools to avoid false positives by using as a fingerprint the key itself. Interestingly, it still provides better time and memory performances than using a standard hash table in the DNA $k$-mer indexing context, with $k=31$, which is a very common value for read comparisons~\cite{Benoit2016}. 

We could improve our technique to recognize key from the original set, using  techniques from the hashing field~\cite{kirsch2006less} or from the set representation field~\cite{belazzougui2013compressed}.
We could thus hope to represent a non-probabilistic dictionary with a tractable memory footprint or achieve a smaller false positive rate with the same or a reduced memory usage.

Regarding the bioinformatics applications for short reads, the results we provide show that alignment-based approaches do not scale when it comes to find similar reads in data sets composed of millions of sequences. The fact that high throughput sequencing data counts rarely less than millions reads justifies our approach based on $k$-mer similarity. Moreover our approach is more straightforward and requires less parameters and heuristics than mapping approaches, that can sometimes turn them into black boxes. However, our alignment-free approach remains less precise than mapping.
An important future work will be to further evaluate our results in terms of sensitivity and precision in comparison to well-known and widely used tool as BLAST~\cite{altschul1990basic}.

As for long reads application, MHAP and GraphMap embed filters to increase their mapping precision, while Minimap and SRC\_linker share the property of integrating no post-treatment heuristics.
We showed we attained other state of the art tools in terms of recall and almost reached their precision without specific developments in our tool. 
We could however increase this precision 
by tuning parameters or by integrating more developments.
We combined high quality results in different ranges of errors with an ability to scale. Such a combination is a strength of our approach in comparison to other tools.
 We shown one particularity of these application of the quasi-dictionary is that they are generic as they can be applied both short and long reads for similar purposes. The simplicity of the tools we presented makes them easy to modify them in any way to better fit to a given problem. Moreover, in their current form they already have straightforward applications examples in biology, such as the building of sequences similarity networks (SSN)~\cite{Atkinsonetal2009} using SRC\_linker. SSN have recently been adapted to address an increasing number of biological questions investigating both patterns and processes: e.g. genomes heterogeneity~\cite{boon2015}; microbial complexity and evolution~\cite{Corel2016}; microbiome adaptation~\cite{Bap2012Clinics,Volkel2016} or to explore the microbial dark matter~\cite{dmSSN}. These applications often bring voluminous reads experiments and large SSN problem instances, where SRC\_linker will scale when other classical tools cannot be applied. Finally, the successful utilization and the perspectives of our structure on bioinformatics problems should not narrow the potential broad applications in other fields, as the general framework could be adapted to many other questions.

\section*{Acknowledgments}
This work was funded by French ANR-12-BS02-0008 Colib'read project. 
We thank the GenOuest BioInformatics Platform that provided the computing resources necessary for benchmarking. We warmly thank Guillaume Rizk and Rayan Chikhi for their work on the MPHF and for their feedback on the preliminary version of this manuscript.

\bibliographystyle{plain}
\bibliography{mybibfile}

\section{Appendix}
In this section, we describe the experiment designed for testing our approach versus alignment-free tools in the recruitment of long and noisy reads.\\
We used MHAP version 2.1.1 with default parameters, Minimap version 0.2-r123 with option \texttt{-Sw 2} and Graphmap version 0.4.1 with default parameters. We tried increasing the recall of Minimap by allowing to index more minimizers using the option  \texttt{-Sw} so that its results would be more comparable to GraphMap's and ours. For MHAP, a similar feature could be tuned too but on our simulations it increased non significantly the recall while decreasing the precision. We chose the best set of parameters ($k=15,  w=2000, s=8$ for 12\% error and $k=15,  w=600,  s=8$ for 15\% error) for SRC\_linker indicated by our simulations.  All tools allowed multi-threading and were launched using 10 threads. We extended the timeout of 5 hours from the short reads experiment regarding the longer sequences to process, reaching 15 hours wallclock.

Summing up, we used the following commands, respectively for MHAP, Minimap, Graphmap and SRC\_linker:
 \begin{verbatim}#java -jar mhap-2.1.1.jar --num-threads 10 -s reads.fa\end{verbatim}
 \begin{verbatim}#minimap -Sw2 -L100 -t10 reads.fa reads.fa\end{verbatim}
 \begin{verbatim}#graphmap owler -t 10 -r reads.fa -d reads.fa -o out.mhap\end{verbatim}
 \begin{verbatim}#short_read_connector.sh -b reads.fa -q fof.txt -t 10  -p out.src 
 -k 15 -w 2000 -s 8\end{verbatim}

We designed a very simple experiment to be able to distinguish true from false similarities  retrieved by the different tools. We chose spots on the caenorhabditis elegans genome (genome version PRJNA13758 from WormBase) separated from hundreds of nucleotides and simulated reads on theses spots of 2K bases long. This creates the following situation. No reads overlap two different spots, and reads share whole-length overlaps. This simple situation enables us to have access to the ground truth about read similarities without using a third-party mapping tool. 
The ground truth is composed of a set of read id couples. Each couple designs two reads simulated from the same locus. Each tested tool also produces a set of read id couples. Recall and precision measures are given by the following formulas: 
\begin{align*}
\mathrm{recall} &= \frac{\text{Number of correctly predicted couples}}{\text{Number of ground truth couples}}\\
 \mathrm{precision} &= \frac{\text{Number of correctly predicted couples}}{\text{Number of predicted couples}}
\end{align*}
The F-measure is provided by the following formula: 
$$\text{F-measure} = 2 \cdot \frac{\mathrm{precision} \cdot \mathrm{recall}}{ \mathrm{precision} + \mathrm{recall}}é$$

We chose C. elegans for its known relative simplicity (it contains few repeats within its DNA sequence), which lowers the chance to mistake a region for another, and we preferred real biological sequence to random sequence to be closer to the biological applications we aim at. Two reads sets of 100K and 1M reads were simulated, using an error profile that mimics PacBio reads \cite{ono2013pbsim} and 12\% error rate, which represents the expected scenario in that sequencing technology.
It is noteworthy that the current sizes of long reads experiments are smaller, but they are expected to grow as sequencing costs decrease.

\end{document}